\documentclass[twocolumn,showpacs,preprintnumbers]{revtex4}
\usepackage{amsmath,amsfonts,amssymb,braket,graphicx}
\usepackage{hyperref}
\usepackage{setspace}
\usepackage[english]{babel}
\usepackage{graphicx}
\usepackage{dcolumn}
\usepackage{bm}
\usepackage{epstopdf}

\def\om {\omega}
\def\be {\begin{equation}}
\def\ee {\end{equation}}
\def\1{\mathchoice{\rm 1\mskip-4.2mu l}{\rm 1\mskip-4.2mu l}{\rm
        1\mskip-4.6mu l}{\rm 1\mskip-5.2mu l}}

\begin{document}

\title{Photon exchange and entanglement formation during the transmission through a rectangular quantum barrier}

\author{Georg Sulyok$^{1}$}
\author{Katharina Durstberger-Rennhofer$^{1}$}
\author{Johann Summhammer $^{1}$}

\affiliation{%
$^1$Institute of Atomic and Subatomic Physics, Vienna University of Technology, 1020 Vienna, Austria}

\date{\today}

\begin{abstract}
When a quantum particle traverses a rectangular potential created by a quantum field both photon exchange and entanglement between particle and field take place. We present analytic results for the transition amplitudes of any possible photon exchange processes for an incoming plane wave and initial Fock, thermal and coherent field states. We show that for coherent field states the entanglement correlates the particle's position to the photon number in the field instead of the particle's energy as usual.
Besides entanglement formation, remarkable differences to the classical field treatment also appear with respect to the symmetry between photon emission and absorption, resonance effects and if the field initially occupies the vacuum state. 
\end{abstract}


\maketitle


\section{Introduction}
\label{sec:intro}
The behaviour of a quantum particle exposed to an oscillating rectangular potential has been studied by several authors under different aspects involving, for example, tunnelling time \cite{Buettiker_Landauer_traversal_time, Stovneng_Hauge}, chaotic signatures \cite{Leonel_barrier_chaos, Henseler_quantum_periodically_driven_scattering}, appearance of Fano resonances \cite{quantum_barrier_Fano_resonances_Lu}, Floquet scattering for strong fields \cite{Reichl_Floquet_strong_fields} and its absence for non-Hermitian potentials \cite{Longhi_oscillating_non-Hermitian_potential}, chiral tunnelling \cite{chiral_tunneling}, charge pumping \cite{charrge_pumping_Wu} and other photon assisted quantum transport phenomena in theory \cite{TienGordon, PAT_Platero_phys_rep, PAT_quantum_transport_Wei} and experiment \cite{PAT_Blick, PAT_Drexler, PAT_Kouwenhoven, PAT_Verghese, PAT_Wyss}. 

In these works, though the potential is treated as a classical quantity, the change of the particle's energy is explicitly attributed to a photon emission or absorption process. Here, we introduce the photon concept in a formally correct way by describing the field generating the potential as quantized. Hence, we pursue the ideas which we started to elaborate in our previous publication \cite{Sulyok_Summi_Rauch_PRA}. There, we only arrived at an algebraic expression for the photon transition amplitudes whereas we now are able to present analytic results for all important initial field states enabling advanced investigations on photon exchange processes and entanglement formation.

In order to compare semiclassical and fully-quantized treatment in our physical scenario, we will at first recapitulate the results of the calculation for a classical field (chap.\ref{sec:classical}). Then, we turn to the quantized field treatment (chap.\ref{sec:quantized}). After presenting the general algebraic solution, we will explicitly evaluate the photon exchange probabilities for an incoming plane wave and for a field being initially in an arbitrary Fock state, a thermal state or a coherent state. The special cases of no initial photons (vacuum state) and of high initial photon numbers will be treated in particular.

\section{Classical treatment of the field}
\label{sec:classical}

\begin{figure}[!htb]
\centering
\includegraphics[width=7cm,keepaspectratio=true]{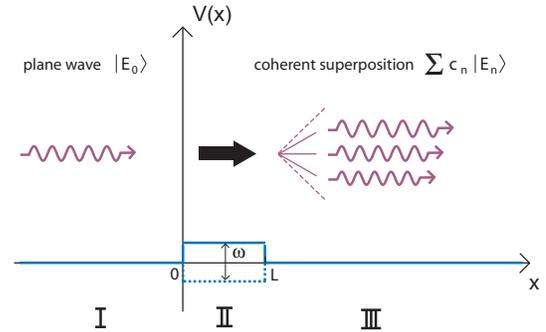}
\caption{Spatial characteristics of the considered potential $V$. It is harmonically oscillating in time with frequency $\omega$ in region II and vanishes elsewhere. An incoming plane wave with energy $E_0\gg V$ is split up into a coherent superposition of plane waves with energy $E_n=E_0+n\hbar \omega$.}
\label{fig:potential2D}
\end{figure}

The potential created by a classical field is a real-valued function of space and time in the particle's Hamiltonian. 
Our considered potential oscillates harmonically in time and is spatially constant for $0\leq x < L$ and vanishes outside.
\begin{equation}
\hat H  = 
\left\{ \begin{array}{ll}
\frac{\hat p^2}{2m}
+ V \cos(\omega t+\varphi), & 
\textrm{if $0 \le x < L $ (region II)}\\
\frac{\hat p^2}{2m}, & 
\textrm{else (region I+III)}
\end{array} \right.
\end{equation}
It therefore corresponds to a harmonically oscillating rectangular potential barrier (see fig.\ref{fig:potential2D}).

The Schr\"odinger equation is solved in each of the three regions separately and then the wave functions are matched by continuity conditions. A general approach based on  Floquet theory \cite{hideo_sambe_floquet} can be found in \cite{Li_Reichl_floquet}. We restrict ourselves to incoming waves whose energy  $E_0$ is much higher than the potential ($E_0 \gg V$). Reflection at the barrier can then be neglected and standard methods for differential equations suffice to find the solution \cite{Summhammer93MultiPhoton, HaavigReifenberger}. If we assume the wave function $\ket{\psi_I}$ in region I to be a plane wave with wave vector $k_0$ we get for the wave function $\ket{\psi_{III}}$ behind the potential barrier
\begin{equation}
\label{eq:classical_solutions}
\ket{\psi_{I}}=\ket{k_0} \, \Longrightarrow \, \ket{\psi_{III}}= \sum_{n=-\infty}^{+\infty} J_n (\beta)\ e^{-i n \eta} \, \ket{k_n}
\end{equation}
where
\begin{eqnarray}
\label{eq:abbreviations_classical1}
\beta &=& 2 \frac{V}{\hbar\omega} \sin\frac{\omega \tau}{2},
\quad \eta=\varphi+\frac{\omega \tau}{2} + \frac{\pi}{2} \\
\label{eq:abbreviations_classical2}
\tau &=& \frac{m L}{\hbar k_0}=\frac{L}{v_0} ,\ k_n^{2}=k_0^2+ \frac{2m}{\hbar} n \omega
\end{eqnarray}
For a more detailed derivation including the solution for region II as well we refer to \cite{Summhammer93MultiPhoton, sulyok_photexchg_pra}.

In summary, a plane wave $\ket{k_0}$ gets split up into a coherent superposition of plane waves $\ket{k_n}$ whose energy is given by the incident energy $E_0$ plus integer multiples of $\hbar \omega$. The transition probability for an energy exchange of $n \hbar \omega$ is just the square of the Bessel function $J_n^{\,2}$ of the $n$-th order. The argument of the Bessel function shows that an increasing amplitude $V$ of the potential also increases the probability for exchanging larger amounts of energy. 

Apart from this expected result, it also exhibits a "resonance"-condition. If the "time-of-flight" $\tau$ through the field region and the oscillation frequency are tuned such that $\omega \tau = 2l\pi,\, l\in\mathbb N$, all Bessel functions $J_n$ with $n \neq 0$ vanish and no energy is transferred at all. The plane wave even passes the potential completely unaltered since $J_0(0)=1$. That's a remarkable difference between an oscillating and a static potential where at least phase factors are always attached to the wave function. An experimental implementation of the classical potential can be found in \cite{sulyok_photexchg_pra,Summhammer95MultiPhotonObservation}.

\section{Quantized treatment of the field}
\label{sec:quantized}

Since the energy exchange between the harmonically oscillating potential and the particle is quantized by integer multiples of $\hbar \omega$ most authors already speak of photon exchange processes although the potential stems from a purely classical field. This notion is problematic since a formally correct introduction of the photon concept requires a quantization of the field generating the potential. For this purpose, the corresponding field equation has to be solved and a canonical quantization condition for Fourier amplitudes of the field is introduced which are then no longer complex-valued coefficients but interpreted as creation and annihilation operators.

For the further, we assume that such a quantum field whose spatial mode is well approximated by the rectangular form generates the potential.  
The quantum system we observe now consists of particle and field together.
The total state $\ket{\Psi}$ of the composite quantum system is an element of the product Hilbert space $\mathcal H_{\rm total}=\mathcal H_{\rm particle}\otimes\mathcal H_{\rm field}$. If the particle is outside the field region the evolution of the state is given by $\hat{H}_0$ composed of the free single-system Hamiltonians $\hat{h}^{\rm p}_0$ and $\hat{h}^{\rm f}_0$ of particle and field 
\begin{eqnarray}
\label{eq:H0}
\hat{H}_0 &=&
\hat{h}^{\rm p}_0 \otimes \1 
+ \1 \otimes \hat{h}^{\rm f}_0 
\\
\label{eq:free_single_hamiltonians}
\hat{h}^{\rm p}_0 &=& \frac{\hat p^2}{2m}, \quad
\hat{h}^{\rm f}_0 = \hbar \om 
\textstyle
\left(\hat a^{\dagger} \hat a + \frac{1}{2}\right)
\end{eqnarray}

\begin{figure}[!htb]
\centering
\includegraphics[width=8cm,keepaspectratio=true]{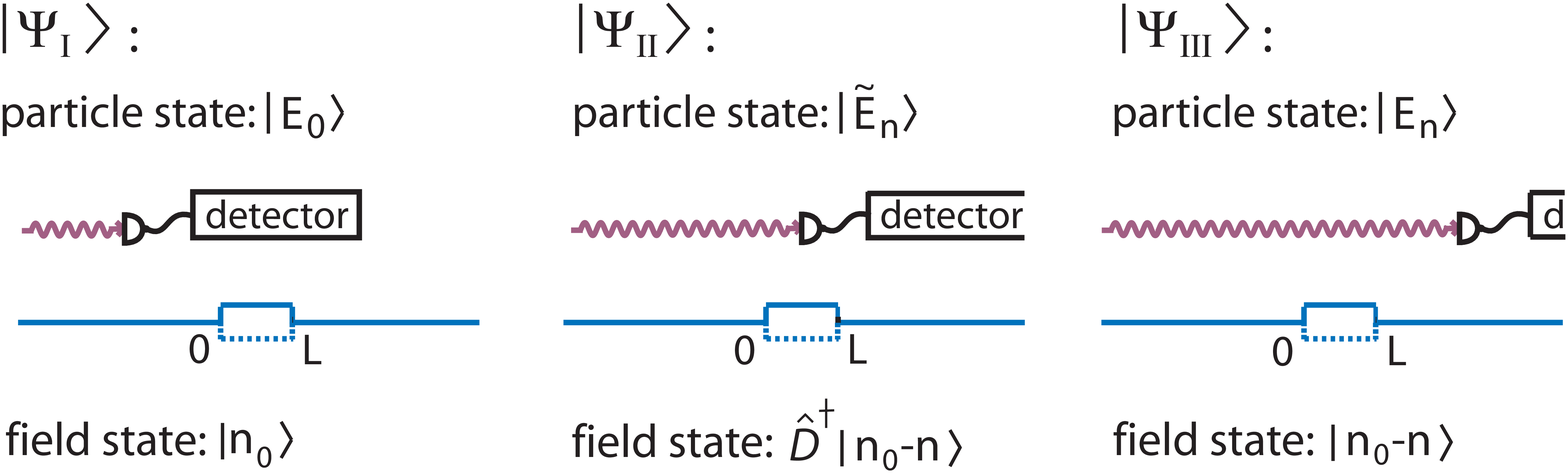}
\caption{In the quantized field treatment, the particle's position determines which of the overall wavefunctions $\ket{\Psi_{I}}$, $\ket{\Psi_{II}}$ or $\ket{\Psi_{III}}$ describes the state of the composite quantum system. The spatial characteristics of the field do not change, it is always present between 0 and L, but the field state changes in accordance with the particle due to their interaction.
\label{fig:quantized_scheme}}
\end{figure}

Interaction between field and particle takes place if the particle is inside the field region, that is, its position coordinate fulfils $0\leq x_{\rm particle}<L$.
Then, the evolution of the composite state $\ket{\Psi}$ is governed by the full Hamiltonian $\hat H = \hat{H}_0 + \hat{H}_{\rm int}$. Basically, the interaction Hamiltonian $\hat{H}_{\rm int}$ is given by the quantized version of the sinusoidal driving term
\begin{equation}
\label{eq:H_int}
\hat{H}_{\rm int} = \lambda \, \1 \otimes \left(\hat a^{\dagger}  + \hat a\right)
\end{equation}
where all constants and the eigenvalue of operator acting on the particle (e.g. spin, charge) have already been absorbed in the coupling parameter $\lambda$. The explicit form of $\hat{H}_{\rm int}$ depends on the actual physical context, for example, dipol interaction for a charged particle in an electromagnetic field or Zeeman-Hamiltonian for uncharged particles in a magnetic field \cite{atom_photon_interactions}. Mind that, although $\hat H_{\rm int}$ in the form of (eq.\ref{eq:H_int}) seems to act solely on the field part of the composite state, the sheer presence of an interaction is connected to the particle's position. Therefore, we again distinguish between three different states $\ket{\Psi_{I}}$, $\ket{\Psi_{II}}$, and $\ket{\Psi_{III}}$ for the composite quantum system (see fig.\ref{fig:quantized_scheme}). 

\subsection{Fock states}
\label{sec:fock}

As in the classical field case, we assume that the kinetic energy of the incoming particle is sufficiently high so that reflection at field entry can be neglected. Then, we can choose as ansatz for $\ket{\Psi_I}$ the particle's state to be a single plane wave with wave vector $k_0$ and the field to be present in a distinct Fock state $n_0$
\begin{equation}
\label{eq:Psi_I}
\ket{\Psi_I}=\ket{k_0} \otimes  \ket{n_0}
\end{equation}
In order to get $\ket{\Psi_{II}}$, we switch to the position space representation of the particle's part of the wave function and match $\ket{\Psi_I}$ at $x_{\rm particle}\equiv x=0$ for all times $t$ with the general solution of the full Hamiltonian $\hat{H}_0 + \hat{H}_{\rm int}$. It is given by an arbitrary linear superposition of plane waves for the particle and displaced Fock states for the field \cite{Sulyok_Summi_Rauch_PRA}. 
The continuity conditions uniquely determine the expansion coefficients and yet  $\ket{\Psi_{II}}$.
At $x=L$, $\ket{\Psi_{II}}$ has to be matched with the general solution of the free Hamiltonian which is given by an arbitrary superposition of plane waves and Fock states. The state $\ket{\Psi_{III}}$ behind the field region then reads
\begin{equation}
\label{eq:final_Fock}
\ket{\Psi_{III}} = \sum_{n=0}^{\infty} t_{n_0 n}\ket{k_{n_0 - n}}\otimes\ket{n},
\qquad k_l^{2}=k_0^2+ \frac{2m}{\hbar} l \omega
\end{equation} 
with
\begin{equation}
\label{eq:transcoef_Fock_algebraic}
t_{n_0 n}=e^{i \bar\lambda^2 \omega \tau} 
\sum_{q=0}^{\infty} 
\bra n \hat D^{\dagger}(\bar \lambda) \ket q 
\bra q \hat D(\bar\lambda) \ket{n_0}
e^{-i (q-n)\omega \tau} 
\end{equation}
where $\hat D$ denotes the displacement operator, $\bar \lambda = \lambda/\hbar\om$ the coupling constant in units of the photon energy, and $\tau=m L/\hbar k_0$ the "time of flight" through the field region as in the classical case (eq.\ref{eq:abbreviations_classical2}).
Details of the calculation as well as the explicit result for $\ket{\Psi_{II}}$ can be found in \cite{Sulyok_Summi_Rauch_PRA}.
The matrix $t_{n_0 n}$ gives the amplitudes for the transition from an initial photon number $n_0$ to the final photon number $n$. The wave vector of the traversing particle changes accordingly from $k_0$ to $k_{n_0-n}$. Every emission of field quanta is absorbed in the kinetic energy of the particle and vice versa.
The final state is the coherent superposition of all such combinations $\ket{k_{n_0-n}}$ and $\ket n$ and therefore highly entangled.

The algebraic form of the transition matrix $t_{n_0n}$ already allows for an intuitive interpretation of the physical processes happening during the transmission. When the particle enters the field the initial Fock state $\ket{n_0}$ experiences a displacement whose amount depends on the coupling constant $\lambda$ (in units of the photon energy). Transitions to other, intermediate Fock states $\ket q$ then occur. When the particle leaves the field a "back-displacement" of the intermediate state takes place. The overlap with the final Fock state $\ket{n}$ at field exit weighted with a phase factor reflecting the energy difference between the intermediate and the final Fock state gives the probability for the transition $n_0 \rightarrow q \rightarrow n$. All intermediate transitions contribute coherently to the transition amplitude $t_{n_0 n}$. 

Summation over all final Fock states $\ket n$ has to be performed in order to receive the total final state $\ket{\Psi_{III}}$. $\ket{\Psi_{III}}$ additionally obtains an overall phase factor from a constant energy shift in region II arising from completing the square in the full Hamiltonian $\hat H$.

\begin{figure}[!htb]
\includegraphics[width=8cm,keepaspectratio=true]{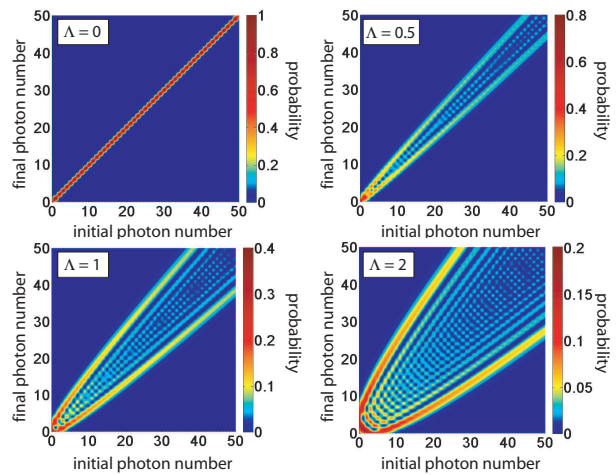}
\caption{Transition probabilities $P_{n_{0},n}$ for initial photon numbers $n_0$ (plotted on the abscissa) and final photon number $n$ (indicated on the ordinata) for coupling strengths $\Lambda = 2\frac{\lambda}{\hbar \omega}\sin\frac{\omega \tau}{2} = 0.0, \Lambda=0.5, \Lambda=1.0, \Lambda=2.0$
\label{fig:fock_x0}}
\end{figure}

The algebraic form of the transition matrix $t_{n_0n}$  (eq.(\ref{eq:transcoef_Fock_algebraic})) can be further developed in order to get an analytic expression. The calculation is straightforward, but rather lengthy and requires the nontrivial Kummer transformation formula for confluent hypergeometric functions. Finally we arrive at
\begin{equation}
\label{eq:transcoef_Fock_analytic}
t_{n_0 n} = e^{i \Phi}
\textstyle
\sqrt{\frac{n_0!}{n!}} 
\,e^{-\frac{\Lambda^2}{2}} \, \Lambda^{n-n_0}
\,\mathcal L_{n_0}^{n-n_0}(\Lambda^2)
\end{equation}
where $\mathcal L_n^{\alpha}(x)$  denotes the generalized Laguerre polynomial and
\begin{eqnarray}
\label{eq:abreviations_quantum}
\textstyle
\Phi&=& \bar\lambda^2 \left(\omega\tau -\sin \omega \tau\right)
 + (n-n_0) \big(\frac{\omega \tau}{2}- \frac{\pi}{2} \big)
\\
\label{eq:coupling_strength}
\Lambda&=&2 \bar\lambda \sin\frac{\om \tau}{2} .
\end{eqnarray}
The coupling strength parameter $\Lambda$ indicates the capacity of the particle-field system to exchange energy and contains the coupling constant $\lambda$ (in units of $\hbar \om$) and the sinusoidal resonance factor that already occurred the classical treatment. The probability that the initial photon number $n_0$ changes to the final photon number $n$ after the transmission of the particle through the field is given by $P_{n_0,n}=|t_{n_0 n}|^2$.
\begin{eqnarray}
 \label{eq:probaility_fock_states}
\textstyle
P_{n_0, n}=\frac{n_0!}{n!} \, e^{-\Lambda^2} \, (\Lambda^2)^{n-n_0} 
\, \big(\mathcal L_{n_0}^{n-n_0}(\Lambda^2)\big)^2 
\end{eqnarray}
In fig.\ref{fig:fock_x0}, the transition probabilities $P_{n_0, n}$ for various coupling strengths $\Lambda$ are depicted.
As in the classical case, the probability for exchanging higher number of photons increases with increasing coupling strength, but absorption and emission of the same number of photons are not equally probable. We have in general $P_{n_0, n}=P_{n, n_0}$ but $P_{n_0, n_0+q}\neq P_{n_0, n_0-q}$.
This asymmetry is reflected in the expectation values of the energy of particle and field after the interaction process. 
\begin{eqnarray}
\label{eq:final_energy_particle}
\bra{\Psi_{III}} \hat h_0^{\rm p} \otimes \1 \ket{\Psi_{III}}
&=&
\frac{\hbar^2 k_0^2}{2m}-\hbar \omega \Lambda^2 \\
\label{eq:final_energy_field}
\bra{\Psi_{III}} \1 \otimes h_0^{\rm f} \ket{\Psi_{III}}
&=&
\textstyle
\hbar \omega \left(n_0+\Lambda^2+ \frac{1}{2} \right)
\end{eqnarray}
Since we assumed a high energetic incoming particle for which reflection could be neglected the net energy transfer goes from particle to field. Not until the initial photon number becomes large with respect to the normed coupling constant $n_0 \gg \bar\lambda$ the symmetry between emission and absorption is restored. We can then use from the appendix of \cite{PolonskiCohenTan}
\begin{equation}
\label{eq:D_fock_large_n0}
\bra{n_0+l}\hat D(\bar\lambda)\ket{n_0+r} = J_{l-r}(2\bar\lambda \sqrt{n_0}) 
,\quad n_0 \gg \bar\lambda
\end{equation}
and apply Graf's addition theorem for Bessel functions in (eq.\ref{eq:transcoef_Fock_algebraic}) to get
\begin{equation}
\label{eq:prob_fock_large_n0}
P_{n_0, n_0+q}=J_q(2 \Lambda \sqrt{n_0} )^2=P_{n_0, n_0-q} 
\end{equation}
Large initial photon numbers indicate the transition to the classical field regime, and indeed, the Bessel function in (eq.\ref{eq:prob_fock_large_n0}) is reminiscent of the classical result (eq.\ref{eq:classical_solutions}). But, if we trace over the field state the particle is still present in an incoherent superposition of the $\ket{k_n}$ weighted with the $J_n^2$ as to be expected from the entangled total state $\ket{\Psi_{III}}$. A proper transition from the quantum to the classical case can only be achieved by starting with a coherent field state (see sec.\ref{sec:coherent}). 

If the length $L$ of the field region and the wave vector $k_0$ are tuned such that the "resonance" condition $\omega\tau= 2\pi n, n\in \mathbb N$ is fulfilled no energy between particle and field is transferred as in the classical case. But, contrary to the classical treatment, an overall phase factor remains in form of $\ket{\Psi_{III}}=e^{i \bar\lambda^2 \omega\tau} \ket{k_0}\otimes\ket{n_0}$ and could be accessible in an interferometric setup.

\subsection{Vacuum state}
\label{sec:vacuum}
Another remarkable feature of the quantum field treatment can be revealed from the investigation of the vacuum state. For a classical field, vacuum is realised by simply setting the potential to zero resulting in an unaltered, free evolution of the plane wave ($\ket{\psi_I}=\ket{\psi_{III}}=\ket{k_0}$). In the quantized treatment, the vacuum is represented by an initial Fock state $\ket{n_0=0}$ which still interacts with the particle and yields as final state $\ket{\Psi_{III}}$ behind the field region
\begin{equation}
\label{eq:state_vacuum}
\ket{\Psi_{I}}=\ket{k_{0}}\otimes\ket{0} 
\quad \Rightarrow \quad
\ket{\Psi_{III}} = \sum_{n=0}^{\infty} t_{0 n}\ket{k_{-n}}\otimes\ket{n}
\end{equation}
with a photon exchange probability
\begin{equation}
P_{0,n}= |t_{0n}|^2= \frac{1}{n!}\, e^{-\Lambda^2}\,\Lambda^{2n}.
\end{equation}
The particle thus transfers energy to the vacuum field leading to a Poissonian distributed final photon number. Let's consider, for example, a superconducting resonant circuit as source of the field. The magnetic field along the axis of a properly shaped coil is well approximated by the rectangular form. A particle with a magnetic dipole moment passing through the coil then interacts with the circuit and excites it with a measurable loss of kinetic energy even if there is classically no field it can couple to. 
The phenomenon that vacuum in quantum field theory does not mean to "no influence" as known from Casimir forces or Lamb shift is clearly visible here as well.


\subsection{Thermal state}
\label{sec:thermal}
\begin{figure}[!htb]
\centering
\includegraphics[width=7cm,keepaspectratio=true]{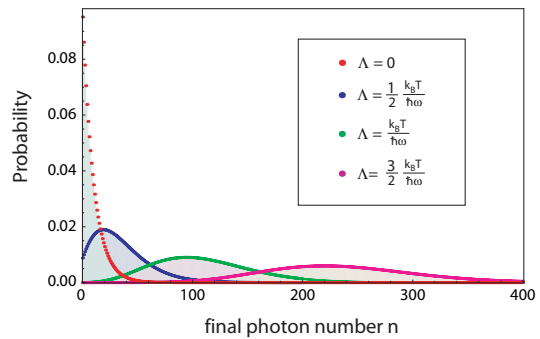}
\caption{Probability distribution of the final photon number for different coupling strengths $\Lambda = 2\frac{\lambda}{\hbar \omega}\sin\frac{\omega \tau}{2}$ if the field was initially in a thermal state (temperature $T, k_B T/\hbar\om \approx 10$).
\label{fig:thermal_figure}}
\end{figure}

In realistic experimental situations, the pure vacuum state can not be achieved. Due to unavoidable coupling to the environment acting as heat bath with a finite temperature $T$ higher photon numbers are excited as well and we encounter the incoherent, so-called thermal state $\rho_{\rm thermal}$ for the field
\begin{equation}
\label{eq:thermal_state}
\rho_{\rm thermal}=
\sum_{n=0}^{\infty} y^n  (1-y) \ket{n}\bra{n}, \qquad
y=e^{-\frac{\hbar\om}{k_B T}}
\end{equation}
We now choose the field to be initially in such a thermal state. After the particle has traversed the field region, the probability $P_n^{\rm therm}$ of finding the field in a distinct Fock state $\ket n$ is given by
\begin{equation}
\label{eq:prob_thermal_state}
\textstyle
P_n^{\rm therm} = e^{-\Lambda^2(1-y)} \ (1-y)\ y^n 
L_n\big(-\frac{\Lambda^2(1-y)^2}{y}\big)
\end{equation}
where $L_n$ denotes the ordinary Laguerre polynomial. As depicted in fig.\ref{fig:thermal_figure}, the initial thermal distribution changes when the coupling strength $\Lambda$ reaches the order of $k_B T/\hbar\om$.

\subsection{Coherent state}
\label{sec:coherent}

\begin{figure}[!htb]
\includegraphics[width=8cm,keepaspectratio=true]{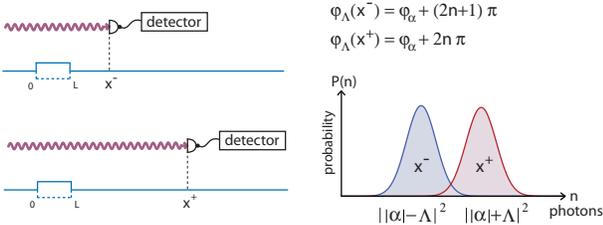}
\caption{Through the phase $\varphi_{\Lambda}(x)$, the final coherent state $\ket{\alpha + \Lambda e^{i\varphi_{\Lambda}(x)}}$ depends on the particle's position. Detecting high (low) photon numbers in the field is therefore correlated to positions $x^+ (x^-)$ and vice versa. 
\label{fig:poissonians}}
\end{figure}

Now, we consider the field to be initially in a coherent state $\ket{\alpha}$ labelled by the complex number $\alpha=|\alpha| e^{i\varphi_{\alpha}}$
\begin{equation}
\label{eq:coherent_state_def}
\ket{\Psi_I}=\ket{k_0}\otimes\ket{\alpha}, \qquad
\ket{\alpha}=e^{-\frac{|\alpha|^2}{2}}\sum_{n=0}^{\infty}\frac{\alpha^n}{ \sqrt{n!}}\ket n .
\end{equation}
For the further evaluation of this expression we start from the algebraic form of the transition matrix (eq.\ref{eq:transcoef_Fock_algebraic}) and work in the position representation of the particle's part of wave function. Expansion of the wave vectors $k_n$ (eq.\ref{eq:final_Fock}) around the initial wave vector $k_0$ enables us to absorb phase factors in the coherent state and evaluate the displacements. The projection onto the position eigenstate $\ket x \in \mathcal H_{\rm particle}$ after the transmission reads 
\begin{eqnarray}
\nonumber
\braket{x|\Psi_{III}} &=& e^{i\bar{\lambda}^2 \om \tau} 
 e^{-i\bar{\lambda}^2 \sin \om\tau} 
 e^{i k_{0} x} \\
&&
\label{eq:psi3_coherent_endres}
e^{ i \Lambda |\alpha| \sin(\varphi_{\Lambda}(x)-\varphi_{\alpha})}
\ket{\alpha + \Lambda e^{i\varphi_{\Lambda}(x)}}
\end{eqnarray}
where
\begin{eqnarray}
\label{eq:psi3_coherent_endres_abbrev}
\varphi_{\Lambda}(x) =\frac{\om \tau}{2} - \frac{\om}{v_0} x - \frac{\pi}{2}
\end{eqnarray}
The entanglement between particle and field is now indicated by the explicit occurrence of the particle's position coordinate $x$ in the final (coherent) field state. If the particle is detected at a certain position $x_1$ the field state is projected onto $\ket{\alpha + \Lambda e^{i\varphi_{\Lambda}(x_1)}}$. We can now place two detectors at positions $x^+$ and $x^-$ which satisfy 
\begin{eqnarray}
\label{eq:phi_plus}
\varphi_{\Lambda}(x^+) &\equiv& \varphi_{\Lambda}^+=\varphi_{\alpha} +2 n \pi \\
\label{eq:phi_minus}
\varphi_{\Lambda}(x^-) &\equiv& \varphi_{\Lambda}^-=\varphi_{\alpha} +2 (m-1) \pi 
\end{eqnarray}
where $n$ and $m$ are arbitrary integers and take a look at the photon number distributions of the related coherent states. The phases $\varphi_{\Lambda}$ are chosen such that the average photon numbers are given by $||\alpha| + \Lambda|^2$ for $x^+$ and $||\alpha| - \Lambda|^2$ for $x^-$ respectively. For a sufficiently high coupling strength $\Lambda \gtrsim \frac{1}{2}$ the corresponding distributions cease to overlap. Detecting the particle around $x^-$ thus increases the probability of having roughly $||\alpha| - \Lambda|^2$ photons in the field whereas detection around $x^+$ is connected to an average photon number of $||\alpha| + \Lambda|^2$. Likewise, finding $||\alpha| + \Lambda|^2$ photons in the field determines the particle's position to be around $x^+$ and analogously for $x^-$ (see fig.\ref{fig:poissonians}). The photon number thus contains information about the particle's position.

\begin{figure}[!htb]
\includegraphics[height=3cm,keepaspectratio=true]{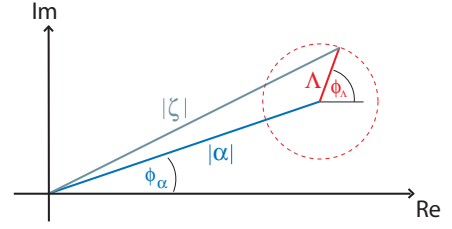}
\caption{For a coherent initial field state $\ket\alpha$ the field state after the transmission is given by an incoherent mixture $\int dx \ \ket\xi\bra\xi$ of all coherent states $\ket\xi =\ket{\alpha + \Lambda e^{i\varphi_{\Lambda}}}$.
\label{fig:coherent}}
\end{figure}

If no measurement on the particle is carried out the field state is obtained from the total density matrix $\rho =\ket{\Psi_{III}}\bra{\Psi_{III}}$ by performing the partial trace over the particle's degrees of freedom. We get an incoherent mixture of coherent states for the field's density matrix
\begin{equation}
\rho_{\rm field} =\int dx 
\ket{\alpha + \Lambda e^{i\varphi_{\Lambda}(x)}}
\bra{\alpha + \Lambda e^{i\varphi_{\Lambda}(x)}}
\end{equation}
which can be illustrated in the Fresnel plane (see fig. \ref{fig:coherent})

Like in case of Fock states, on average, the particle transfers energy to the field as indicated by the expectation values
\begin{eqnarray}
\label{eq:final_energy_particle_coherent}
\bra{\Psi_{III}} \hat h_0^{\rm p} \otimes \1 \ket{\Psi_{III}}
&=&
\frac{\hbar^2 k_0}{2m}-\hbar \omega \Lambda^2 
\\
\label{eq:final_energy_field_coherent}
\bra{\Psi_{III}} \1 \otimes \hat h_0^{\rm p} \ket{\Psi_{III}}
&=&
\textstyle
\hbar \omega \left(|\alpha|^2 + \Lambda^2 + \frac{1}{2} \right)
\end{eqnarray}

If we increase the mean photon number such that we can neglect the coupling strength $\Lambda$ against $|\alpha|$ we can simplify (eq.\ref{eq:psi3_coherent_endres}) and arrive at
\begin{equation}
\ket{\Psi_{III}}= e^{i\bar{\lambda}^2 \omega \tau} e^{-i\bar{\lambda}^2 \sin\omega \tau}\sum_{n=-\infty}^{+\infty} J_n(\Lambda |\alpha|) e^{-i n \eta} \ket{k_n}\otimes
\ket{\alpha}\label{eq:psi3_coherent_high_phot_endres}
\end{equation}
where we have use the abbreviation $\eta$ of the classical section (eq.\ref{eq:abbreviations_classical1}) with $\varphi_{\alpha}\ \widehat{=}-\varphi$. 
Disregarding the back action of the particle on the field thus leads to a simple product state of the composite quantum system and therefore to disentanglement. By tracing over the field, we obtain the particle's state which is now a coherent superposition of $\ket{k_n}$ weighted with the Bessel functions $J_n$ and a phase factor $e^{-in\eta}$ as in the classical case. A general survey on the correspondence between time-independent Schr\"odinger equations for the composite particle-field system and time-dependent Schr\"odinger equations for the particle alone that contain the expression for the classical field as potential term can be found in \cite{Braun_Briggs_Classical_limit}.

If we choose the initial coherent state $\ket\alpha$ to be the vacuum state $\ket 0$ and therefore set $\alpha=0$ in (eq.\ref{eq:psi3_coherent_endres}) we consistently end up with the same final state as in (eq.\ref{eq:state_vacuum}). 

At resonance ($\omega\tau= 2\pi n, n\in \mathbb N$), no photon exchange takes place and the initial state again only obtains an overall phase factor and becomes $\ket{\Psi_{III}}=e^{i \bar\lambda^2 \omega\tau} \ket{k_0}\otimes\ket{\alpha}$ after the interaction.

\section{Conclusion}
\label{sec:conclusion}
The quantum mechanical scattering on a rectangular potential created by a quantum field is completely analytically solvable for incoming particles whose energy is high enough to neglect reflections. Transition amplitudes and photon exchange probabilities can be entirely expressed in terms of standard functions for the most important types of initial field states, that is, Fock, thermal, and coherent states.  
The quantized treatment of both particle and field reveals their entanglement in the interaction process. Therefore, the setup could be of interest for quantum information experiments where a spatially fixed (field) and a moveable component (particle) are required. For Fock states, entanglement actually occurs between the energy eigenstates of the particle and the photon number states of the field, but, for a coherent initial field state, the particle's position and the photon number get entangled. 

The Schr\"odinger equation of the composite system is time-independent and thus, the total energy is conserved in the transmission process. Though, photon emission and absorption are generally not equally probable, on average, the high-energetic, incoming particle transfers energy to the field. Only if the photon number in the field becomes large, the symmetry between emission and absorption is restored. However, in case of pure Fock states, entanglement is nevertheless maintained and the energy transfer happens incoherently. Just for coherent field states whose mean photon number is high against the coupling strength so that the influence of the particle on the field can be neglected the transition to the classical, coherent energy exchange becomes visible. 

A remarkable feature of the fully quantized treatment is the interaction with the vacuum. Though from the classical point of view a free evolution of the particle should take place, the particle transfers energy to the field and their combined state changes. 

For the experimentally more realistic situation of not a pure vacuum but a thermal field state visible effects occur once the coupling constant becomes comparable to the thermal energy ($k_B T$) of the environmental heat bath.  

At resonance, that is when the length of the field region and the particle's wavelength are related such that destructive interference suppresses any photon exchange, the wave function nevertheless changes and obtains an overall phase factor. In the quantized treatment, a completely unaltered evolution only happens in the trivial case of a vanishing coupling constant.

\section{Acknowledgements}
This work was supported by the Austrian science fund (FWF) projects T-389 and P-24973.

\end{document}